\documentclass[a4paper,12pt]{article}

\usepackage{epsfig,float}
\usepackage[left=2cm,right=2cm]{geometry}
\usepackage{a4wide,graphicx}

\usepackage{graphicx,amsmath,amssymb,bbm, ulem}
\usepackage[square,numbers,sort&compress]{natbib}
\usepackage[utf8]{inputenc}
\usepackage[english]{babel}
 \usepackage{authblk}
\newcommand{\be}{\begin{equation}}
\newcommand{\ee}{\end{equation}}
\newcommand{\ba}{\begin{eqnarray}}
\newcommand{\ea}{\end{eqnarray}}

\title{Light- and strange-quark mass dependence of the $\rho(770)$ meson properties}

\author[1]{R. Molina}
\author[2]{J. Ruiz de Elvira}
\affil[1]{Departamento de F\'isica Te\'orica II, Plaza Ciencias, 1, 28040 Madrid, Spain \& Institute of Physics of the University of S\~ao Paulo, Rua do Mat\~ao, 1371 -Butant\~a, S\~ao Paulo -SP, 05508-090}
\affil[2]{Albert Einstein Center for Fundamental Physics, Institute for Theoretical Physics,
University of Bern, Sidlerstrasse 5, 3012 Bern, Switzerland}

\date{ }

\begin{document}

\maketitle

\begin{abstract}
From an analysis of recent ($I=J=1$)-$\pi\pi$-phase-shift and pseudoscalar-meson decay-constant lattice data on two distinct chiral trajectories, where either the sum of the up, down and strange quark masses, or the mass of the strange quark is kept fixed, we extract the light and strange quark mass dependence of the rho meson parameters, and make predictions of those on chiral trajectories which involve lighter masses than the 
physical strange quark mass. We find that the mass of the rho meson can get as light as $700$ MeV for strange quark mass zero at physical pion masses. While the ratio of the couplings to the $\pi\pi$ and $K\bar{K}$ channels is equal to $\sqrt{2}$ at the SU(3) symmetric chiral trajectory.\footnote{Talk given by R. Molina at the Workshop BLED 2019.}
\end{abstract}



\section{Introduction}
 In the past, most  of the LatticeQCD simulations have been done on chiral trajectories $m_s=m_s^0$\footnote{``0'' means the physical value.}. Therefore, previous analysis of lattice data cannot track the behavior of pseudoscalar decay constants and rho meson parameters on trajectories involving variations of the strange quark mass. Recently, the CLS Collaboration has generated ensembles on chiral trajectories like TrM$=c$ (with $c=m_u+m_d+m_s$) in large volumes \cite{Bruno:2016plf,Andersen:2018mau}. Thus, the hadron properties in these trajectories will manifest as a consequence of both, variations in the light and strange quark masses. In this talk, I present the results of a global analysis of lattice data over the TrM$=c$ and $m_s=m_s^0$ trajectories, which include the pseudoscalar decay constant data from Refs. \cite{Bruno:2016plf,Blum:2014tka,Bazavov:2010hj,Bazavov:2009bb,Aubin:2008ie} and phase shift data of Ref. \cite{Dudek:2012xn,Wilson:2015dqa,Bulava:2016mks,Andersen:2018mau}. 
 \section{Theoretical Framework}
 To perform the lattice data analysis, we employ the inverse amplitude method \cite{Truong:1988zp} based on one-loop Chiral Perturbation Theory (NLO ChPT) \cite{Gasser:1983yg,Gasser:1984gg} taking the expressions of Ref.  \cite{GomezNicola:2001as} similarly as in Ref. \cite{Nebreda:2010wv} for the scattering amplitudes $t_2$ and $t_4$ (${\cal O} (p^2)$ and ${\cal O} (p^4)$) of the $\pi\pi-K\bar{K}$ coupled-channel system.
The scattering amplitude, $t$, read as
\be
t=t_2[t_2-t_4]^{-1}t_2\ .\label{eq:iam}
\ee
The elements $t_{ij}$ related to the channels $i,j$ of the 2x2 $t$-matrix are related to $S$-matrix elements, $S_{ij}=\delta_{ij}+2\, i\sqrt{\sigma_i\sigma_j}\,t_{ij}$, where
$\sigma_i =\theta( \sqrt{s}-2\,m_i) \sqrt{1-4\,m_i^2/s}$, and the $S$-matrix is parameterized as $S_{ii}=\eta e^{2i\delta_i}$, $i=1,2$; and $S_{21}=S_{12}=i(1-\eta^2)^{1/2}e^{i(\delta_1+\delta_2)}$.
 The relations for the pseudoscalar meson masses and decay constants of NLO ChPT \cite{Gasser:1983yg,Gasser:1984gg} are used. The chiral trajectories followed by the NLO masses, $M^2_K(M^2_\pi)$ are determined from their LO relations. The ones considered here are
\begin{eqnarray}
M_{0K}^2= \left\{
{\renewcommand{\arraystretch}{2}
     \begin{array}{lr}
      -\frac{1}{2}M^2_{0\pi}+c\,B_0;&Tr M =c\\
      +\frac{1}{2}M^2_{0\pi}+k\,B_0;& m_s=k\\
      M^2_{0\pi};&m_s=m_{ud}\\ 
     (r+m_s)\,B_0;\,M_{0\pi}^2=2rB_0& m_{ud}=r\ ,\\
     \end{array}}
   \right.\nonumber\\
\end{eqnarray}
and $m_\pi=m_\pi^0$. In the above relations, $B_0=\frac{\Sigma_0}{f_0^2}$, with $\Sigma_0=-\langle 0 \vert\bar{q}q \vert 0\rangle_0$, $m_{ud}=(m_u+m_d)/2$, and $f_0$ stands for the pion decay constant in the chiral limit.
The free parameters are the NLO LECs, $L_{12}^r=2\,L_1^r-L_2^r$ and $L_i^r$, $i=3,8$, and $\lbrace c,k\rbrace\times B_0$, which are adjusted to the chiral trajectories. In the fits, $\mu$ is fixed to $770$ MeV, and $f_0$ is set to $80$ MeV.
 Pseudoscalar meson decay constant data and $\rho$-phase-shift data are fitted simultaneously.
The function minimized is the $\chi^2$, defined as,
\ba
\chi^2=(\vec{W}_1-\vec{W}_0)^TC^{-1}(\vec{W}_1-\vec{W}_0)+\sum_{ij}(h_{ij}-h_{ij}^l)^2/e^{l\,2}_{ij}+\lambda\sum_{ij}\int  \vert (S\,S^\dagger)_{ij}-\delta_{ij}\vert^2\, dE\nonumber\\\label{eq:chi2}\ea
where $\vec{W}_0$ is the vector of eigenenergies measured on the lattice, $C$ the covariance matrix of these energies, and $\vec{W}_1$ the corresponding energies of the fit function. Instead of fitting directly the eigenenergies from lattice, these are reconstructed by means of a Taylor expansion, and phase shift data are fitted as done first in Refs. \cite{Hu:2016shf,Hu:2017wli}. This avoids the discretization of loops. In the above equation, $i=1,3$ and $j=1,n$, with $n$ the number of data. $h_1=m_\pi/f_\pi$, $h_2=m_K/f_K$ and $h_3=m_K/f_\pi$. The superscript $l$ indicates values of these ratios from lattice simulations.
 The last term in Eq. (\ref{eq:chi2}) is added to guarantee that the LECs obtained satisfy unitarity at some degree depending on the $\lambda$ value. We checked that for $\lambda\simeq 40$, the values of the LECs are stable, the minimum value of $\chi^2(\lambda)$ lies in a flat range of $\lambda$, and the S-matrix obtained is unitary at a higher degree.
The bootstrap method is used to evaluate the errors assuming that the lattice energies are multivariate normally distributed with the same original covariance matrix, and the resampled-phase-shift data are obtained as a function of the energy expanded at linear order. Additionally, we assume that the lattice spacing in Ref. \cite{Bruno:2016plf}, where two sets of lattice spacings are determined in different ways, is normally distributed with the mean the average between the two different determinations and the typical deviation half the difference between them. 
Results are shown with error bands that mean $68$ and $95$ \% confidence intervals (CI) evaluated from the corresponding quantiles of phase shifts and decay constant ratios. 
\section{Results}
\begin{figure*}
\centering
\includegraphics[scale=0.45]{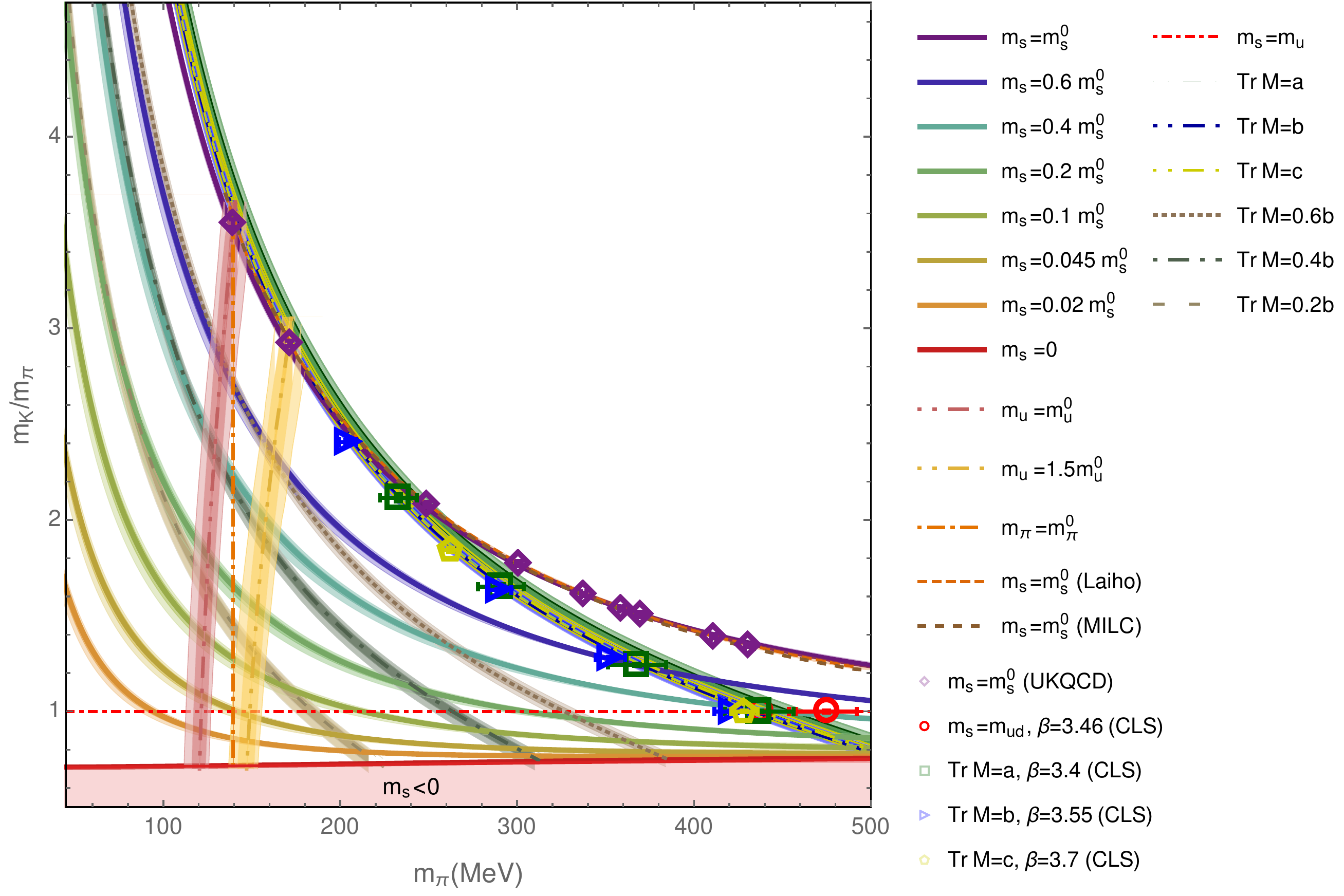}\\
\caption{Chiral trajectories in comparison with lattice data.}
\label{fig:mkmpi}
\end{figure*}

\begin{figure*}
\centering
\includegraphics[scale=0.45]{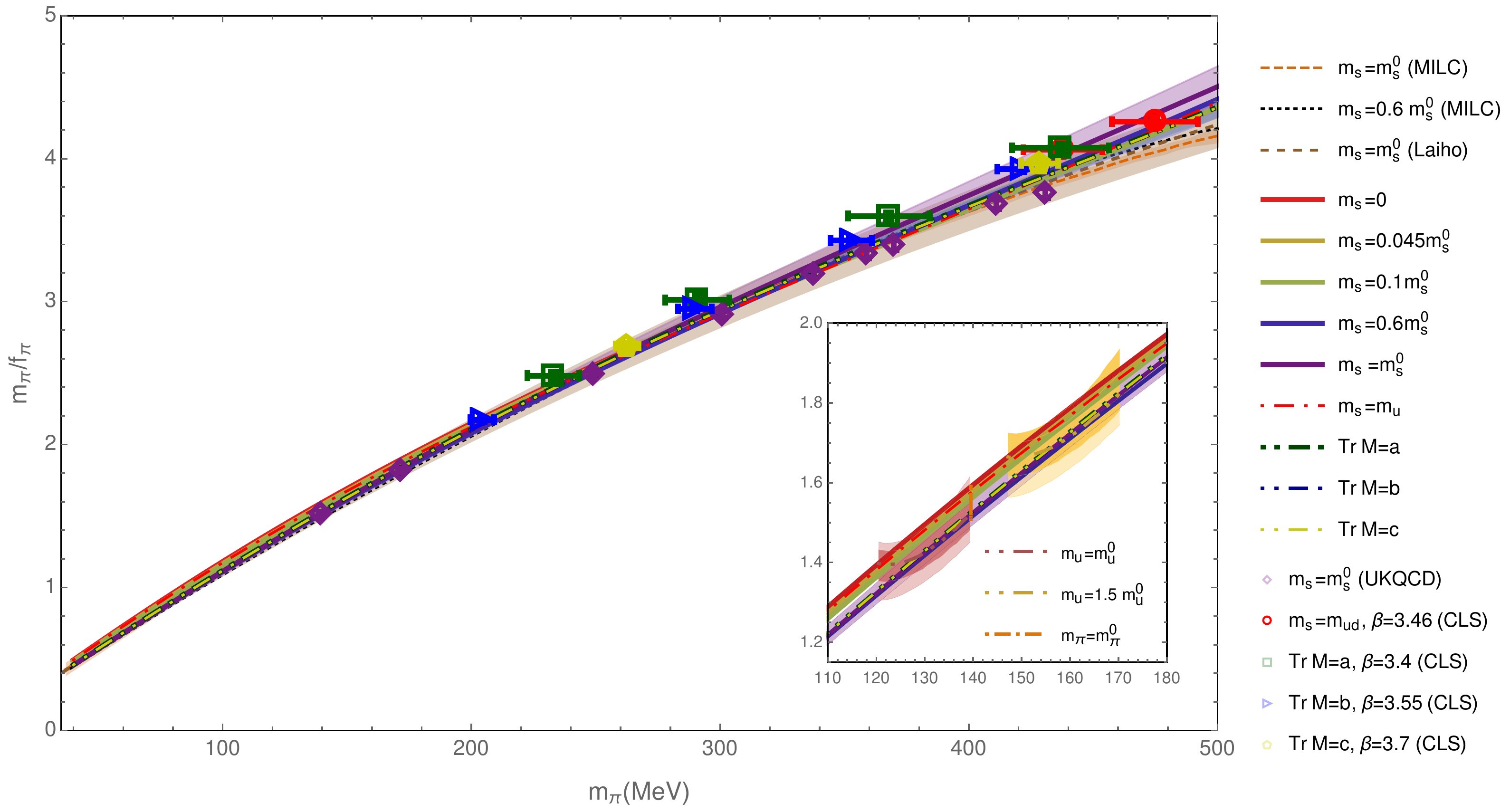}\\
\caption{The ratio $m_\pi/f_\pi$ in comparison with lattice data.}
\label{fig:mpifpi}
\end{figure*}

\begin{figure*}
\begin{center}
\begin{tabular}{cc}
 \hspace{-0.8cm}\includegraphics[scale=0.34]{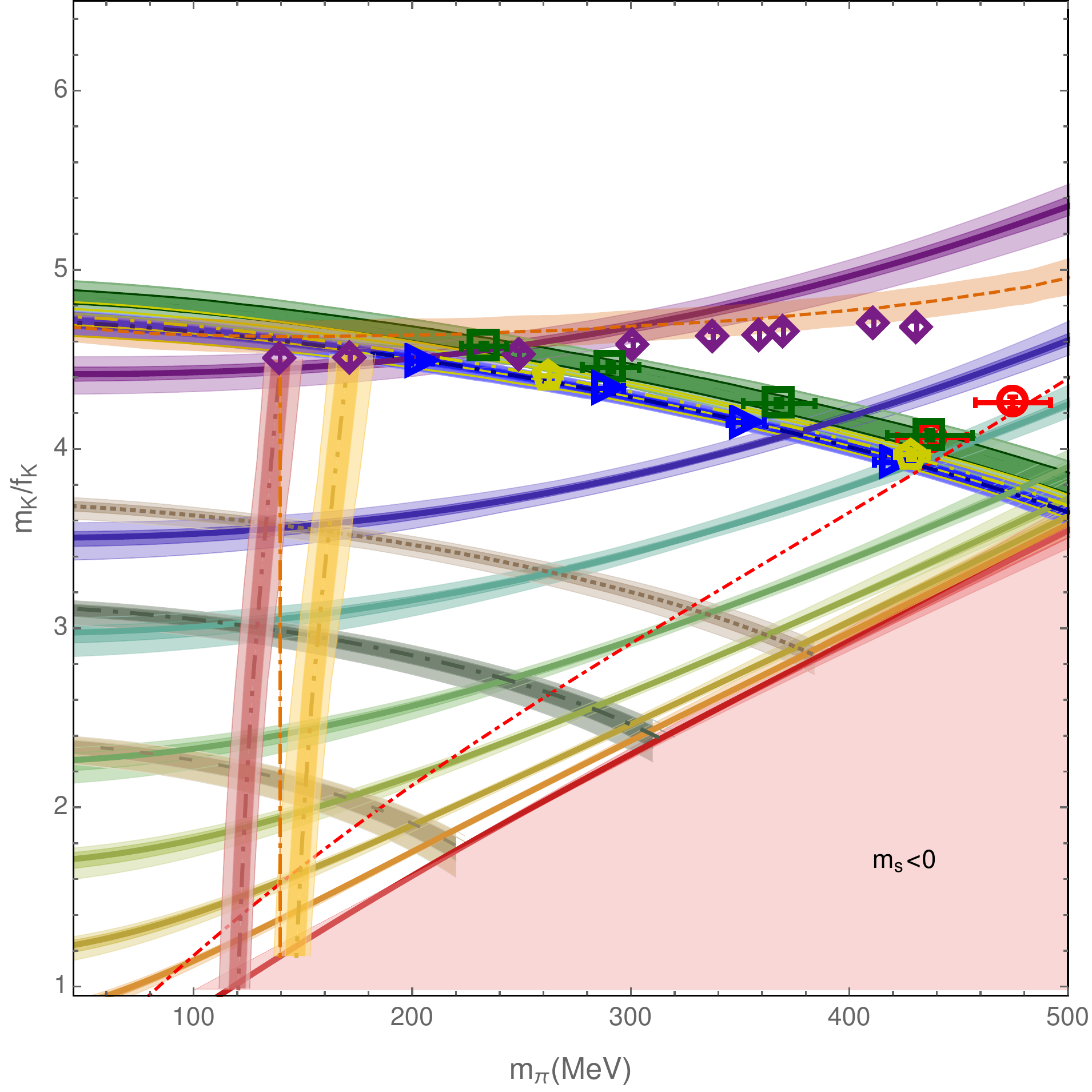} &\includegraphics[scale=0.34]{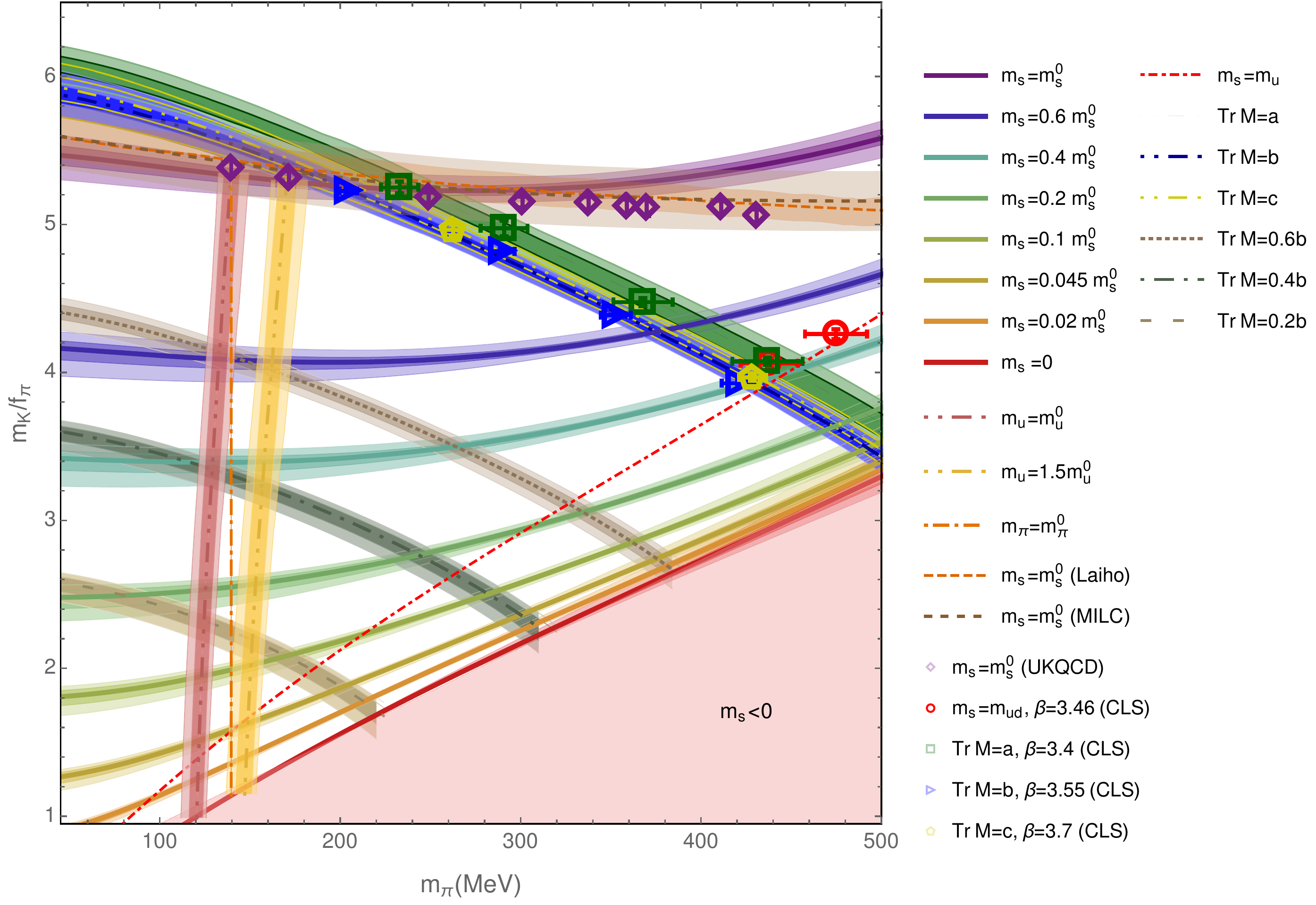}\\
 \end{tabular}
 \end{center}
\caption{Decay constant ratios, $m_K/f_\pi$ and $m_K/f_K$ in comparison with lattice data.}
\label{fig:mkfk}
\end{figure*}

\begin{figure*}
\begin{center}
\begin{tabular}{cc}
 \includegraphics[scale=0.4]{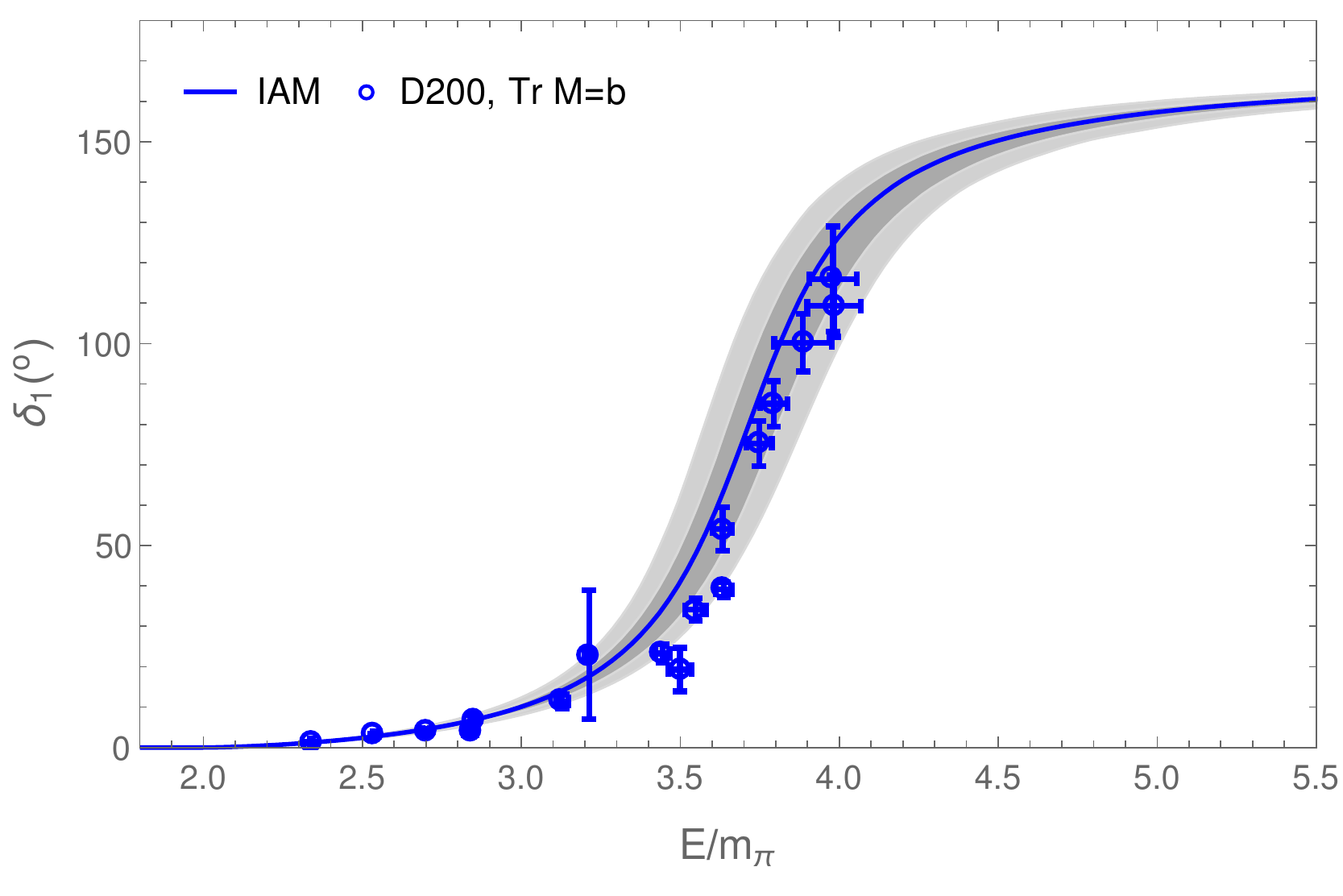} &\includegraphics[scale=0.4]{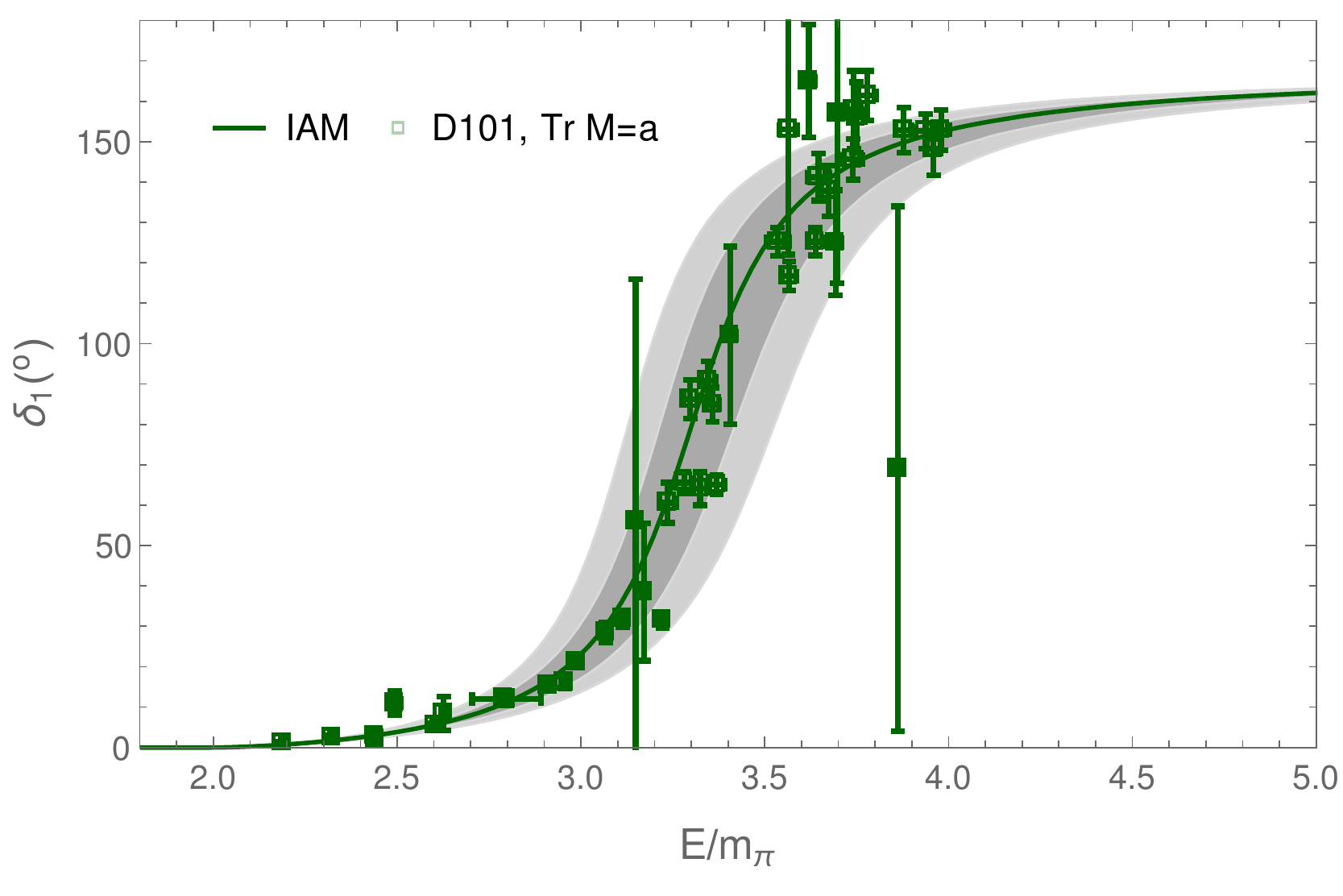}\\
  \includegraphics[scale=0.4]{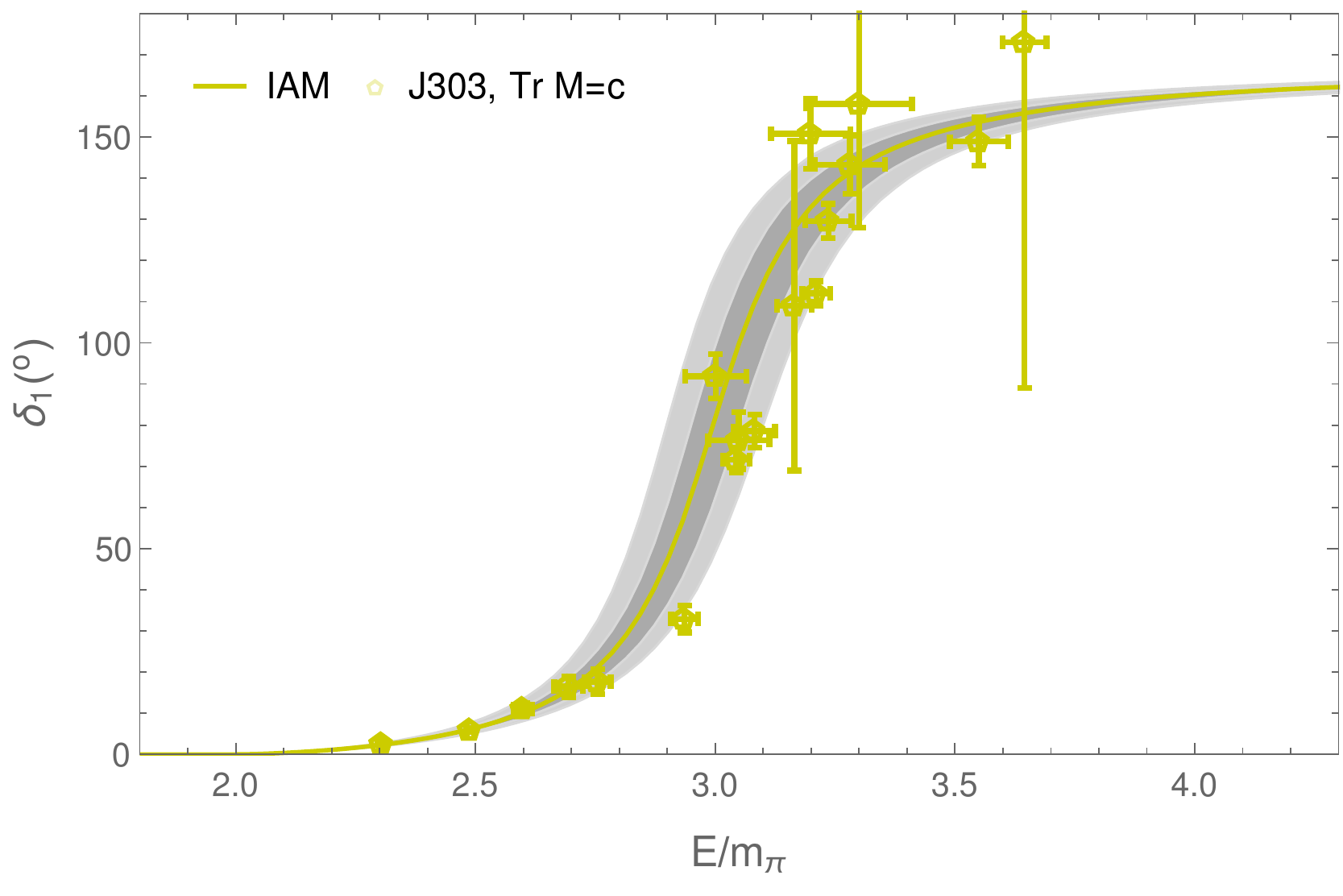}&\includegraphics[scale=0.4]{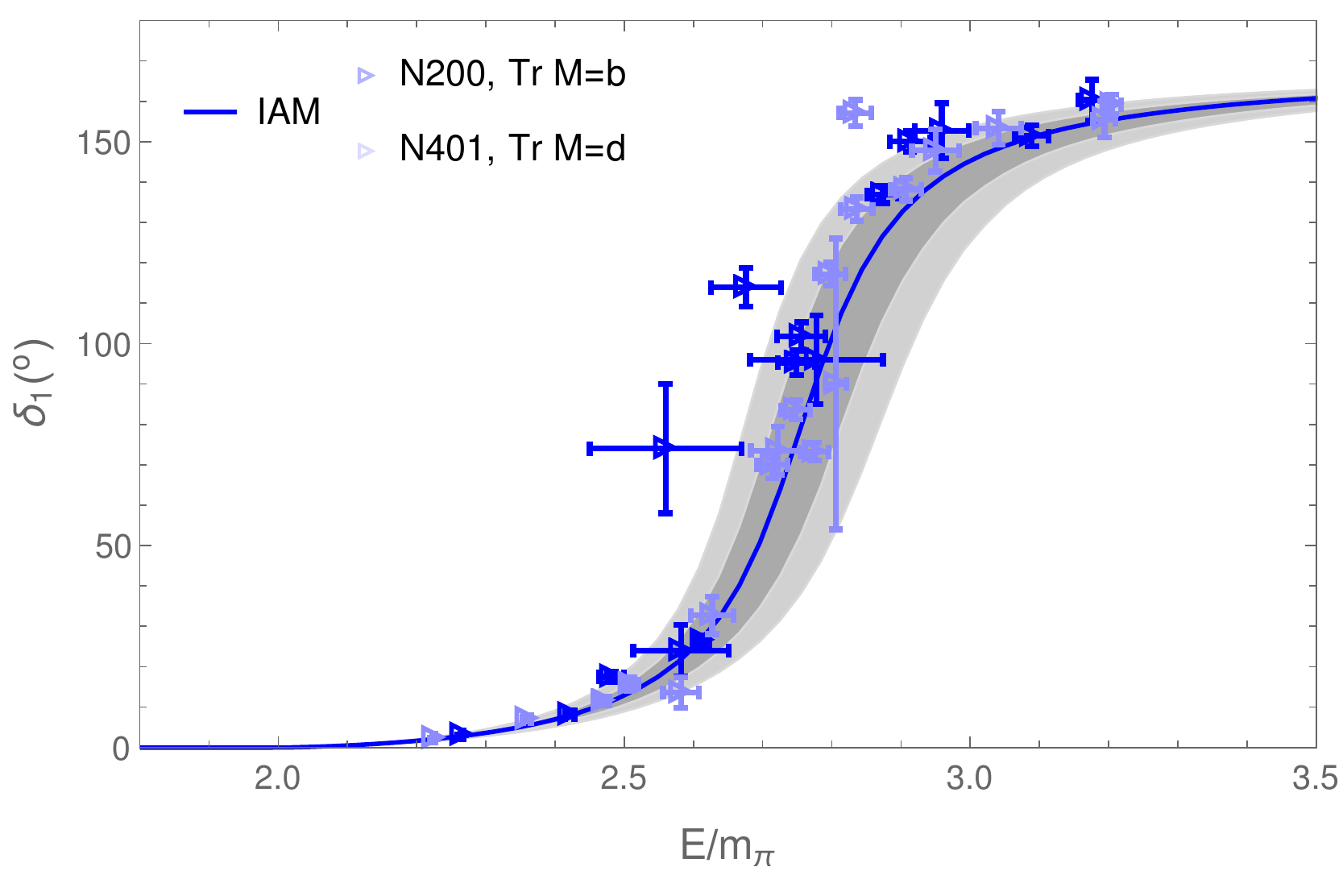}\\
  \includegraphics[scale=0.4]{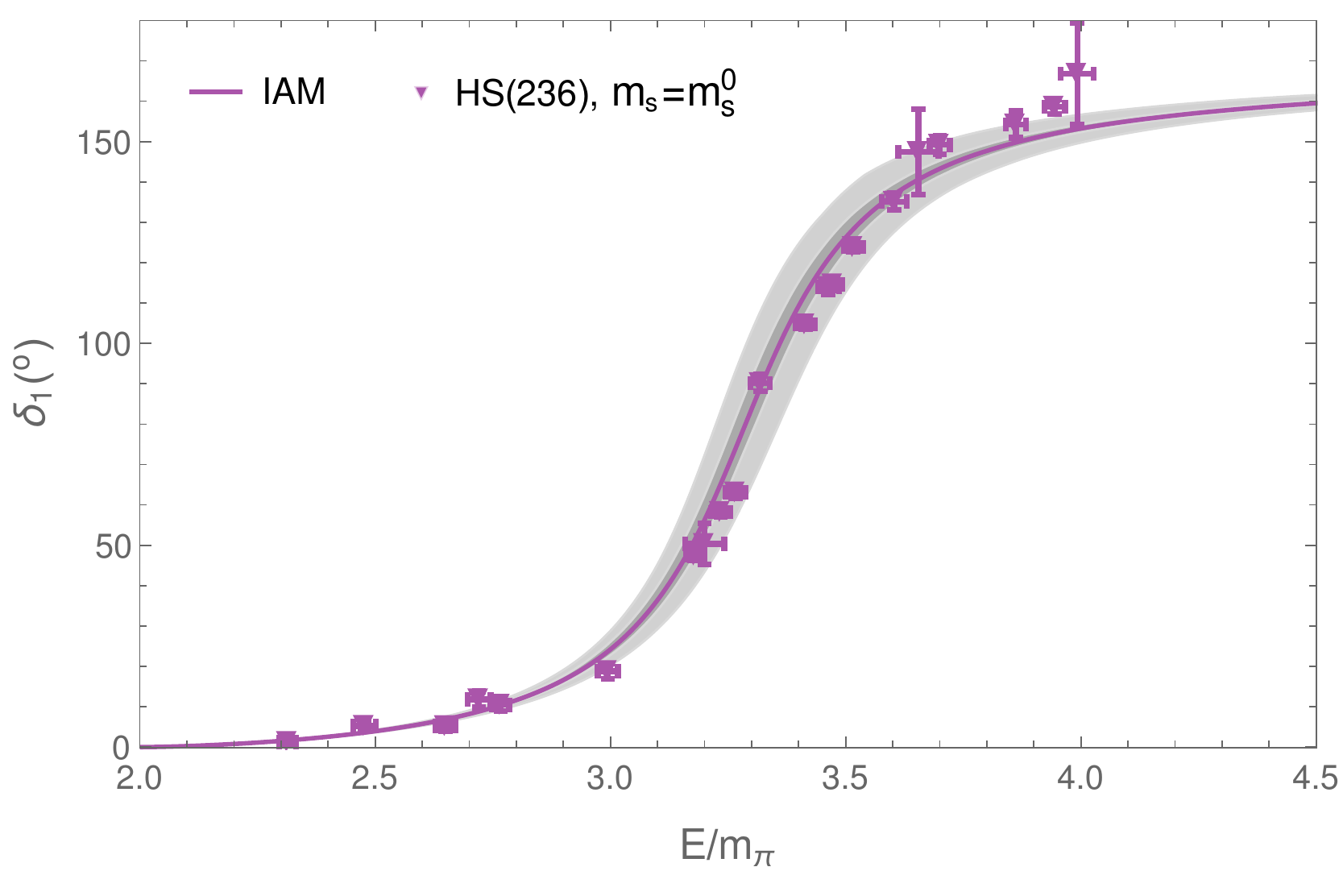} &\includegraphics[scale=0.4]{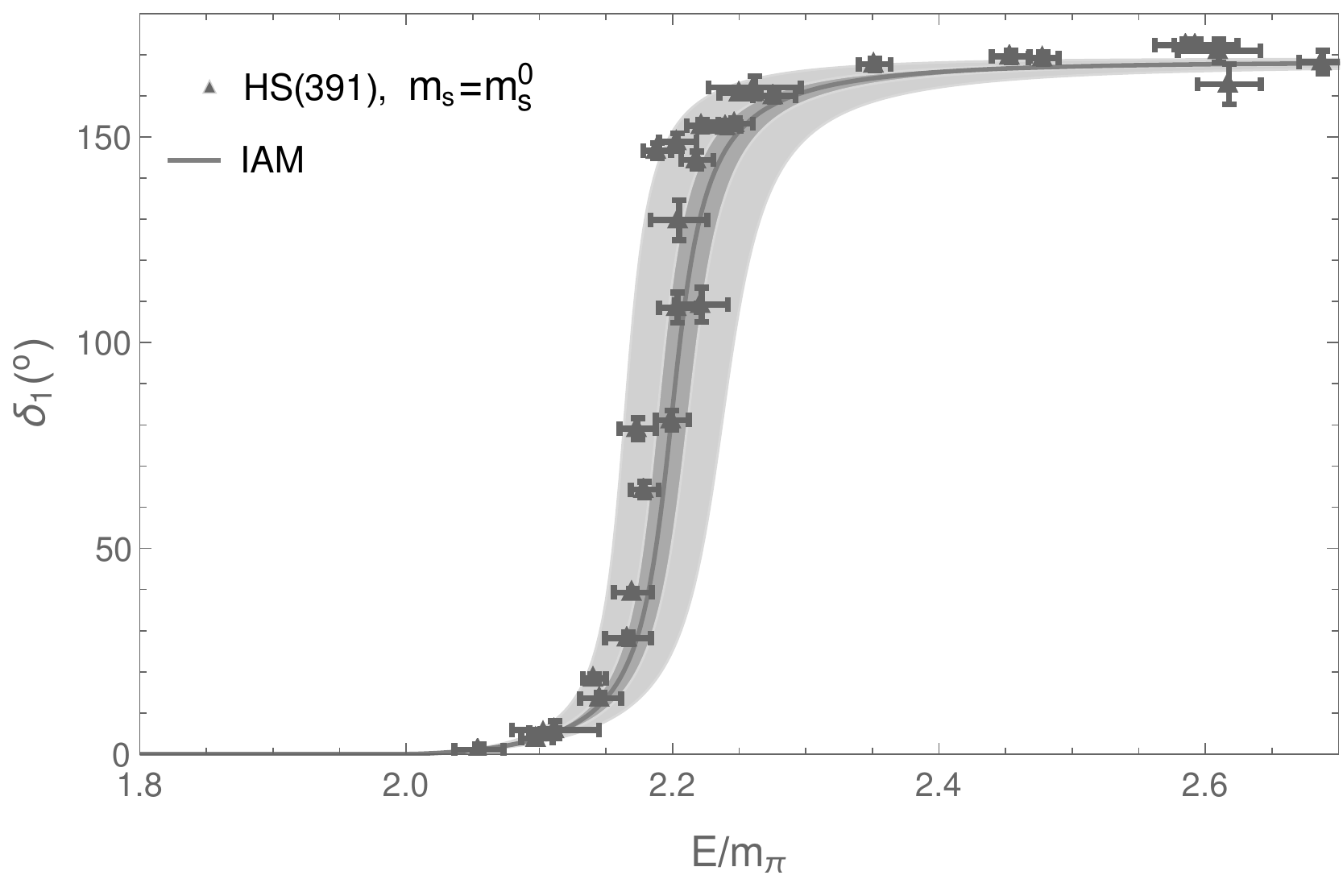}\\
  \includegraphics[scale=0.4]{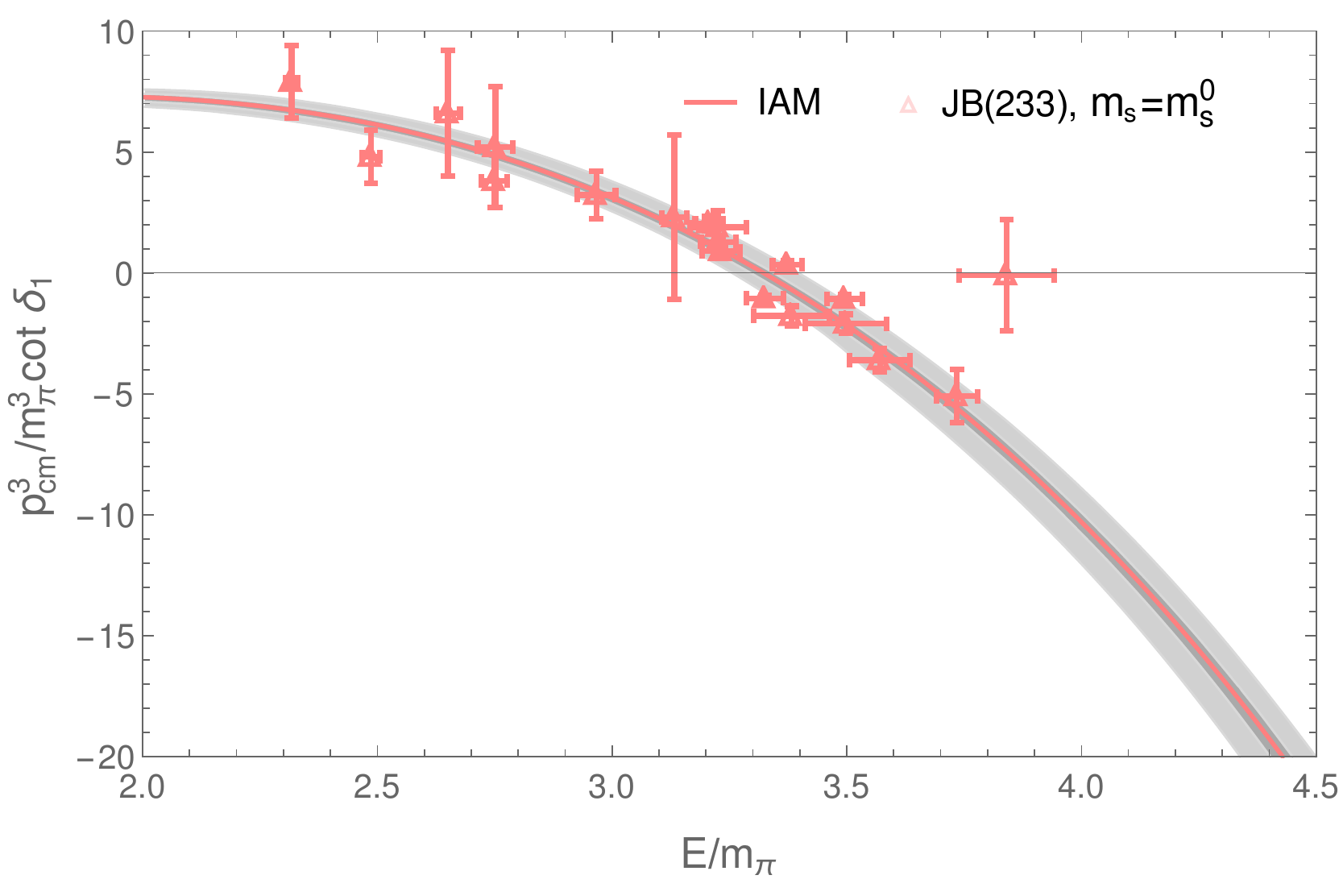} &\includegraphics[scale=0.4]{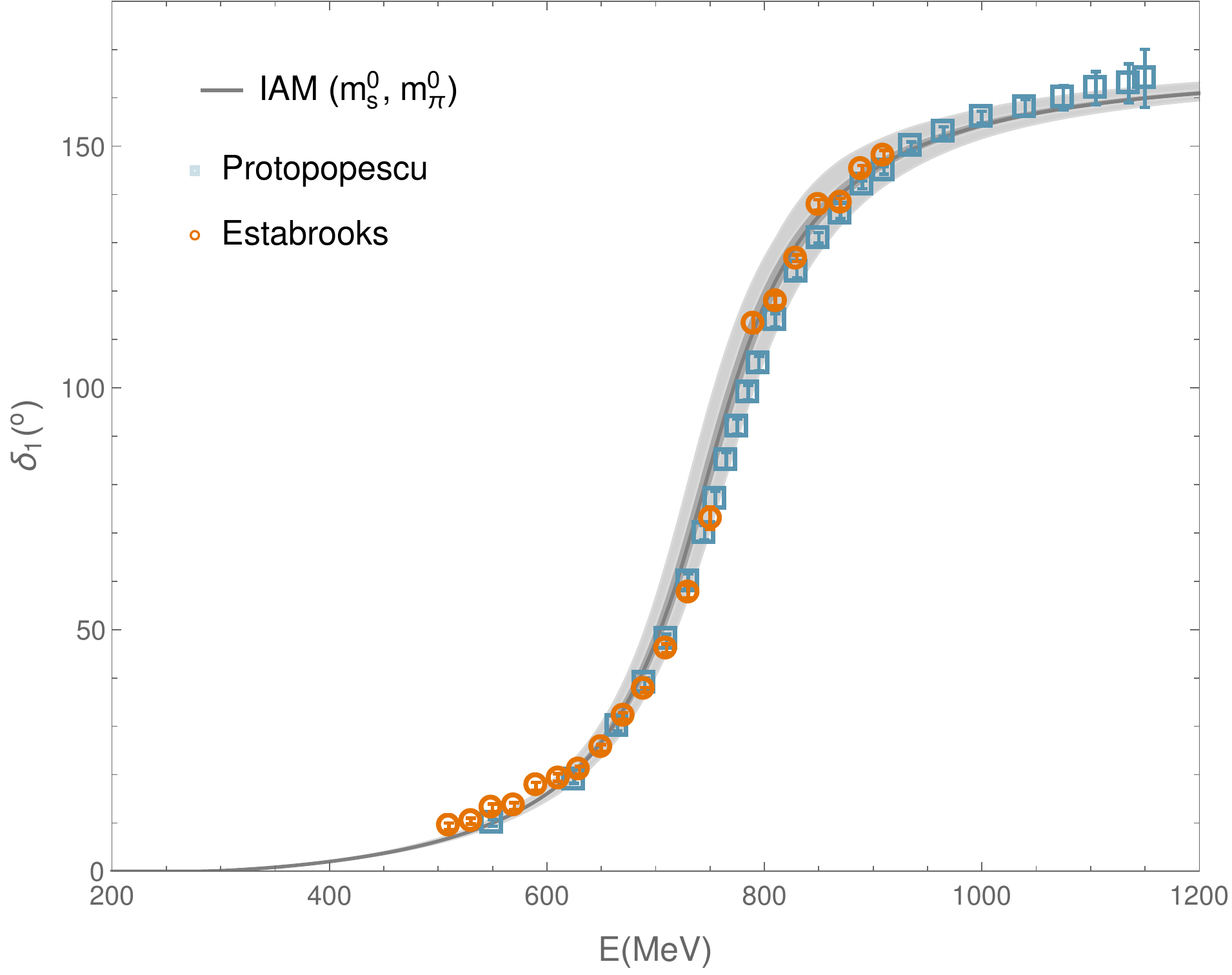}\\
 \end{tabular}
 \end{center}
\caption{Result of the fit in comparison with the TrM$=c$ data of the CLS ensembles, D200, D101, J303, and N200, N401, for pion masses $m_\pi\simeq 200,230,260$ and $290$ MeV, respectively, in the first and second row. In the third row, the result for the $\rho$ phase shifts in the $m_s=m_s^0$ trajectory, in comparison with the HS (HadSpec) data at $m_\pi=236$ and $391$ MeV, and in the fourth row, the result of the fit in comparison with the data of Ref. \cite{Bulava:2016mks} (JB) for $m_\pi=233$ (left), and extrapolation to the physical point in comparison with the experimental data (right).}
\label{fig:phase}
\end{figure*}

In Figs. \ref{fig:mkmpi}, \ref{fig:mpifpi}, and \ref{fig:mkfk}, the chiral trajectories studied together with the decay constant ratios are plotted. The lattice data fitted correspond to the extrapolation to the continuum limit with finite volume effects corrected. For $m_s=m_s^0$, these are, UKQCD\cite{Blum:2014tka} (purple diamonds), MILC\cite{Bazavov:2010hj,Bazavov:2009bb}   (brown dashed curves with light-brown error bands\footnote{The error band for the MILC data is extrapolated from the error at the physical point.} and Laiho\cite{Aubin:2008ie} (orange dashed curves and error bands). For other trajectories $m_s=k$, there are no much data as commented previously, except for the ratio $m_\pi/f_\pi$ extracted by MILC\cite{Bazavov:2010hj} (brown dotted line). The TrM$=c$ data from the CLS Collaboration are given for different strengths of a parameter of the lattice simulation, $\beta=3.4$ (green square), $3.55$ (blue triangle), and $3.7$ (yellow pentagon). The error in the x-axes correspond to the half the difference between the two different lattice spacing determinations. The other trajectories plotted are $m_s=\lbrace0,0.02,0.045, 0.1,0.2,0.4,0.6\rbrace\,m_s^0$, and $m_u=\lbrace 1,1.5\rbrace \,m^0_u$, $m_\pi=m_\pi^0$, which start near $m_s\simeq0$, cross the symmetric line, and end up at the $m_s=m^0_s$ curve.
As seen, all ratios and chiral trajectories are reproduced well inside the $95$ \% CI till $m_\pi\simeq 400$ MeV, where the ChPT predictions start to deviate. 
Phase shifts are very well described being also inside the $95$ \% CI, as shown in Fig. \ref{fig:phase}. The extrapolation to the physical point in comparison with experimental data is plotted in Fig. \ref{fig:phase} (bottom-right). The agreement with the experimental data is impressive. The physical point we get for the masses and decay constant ratios is given in Table 
\ref{tab:physdecayra}. The values of the LECs are given in Table \ref{tab:lecsgl}. For $L_4^r,L_5^r,L_6^r$ and $L_8^r$, we obtain values in line with the Flag average \cite{Aoki:2019cca}. However, notice that our values are much more precise since our result comes from an analysis of data over several chiral trajectories.

\begin{figure*}
\begin{center}
\begin{tabular}{cc}
\includegraphics[scale=0.3]{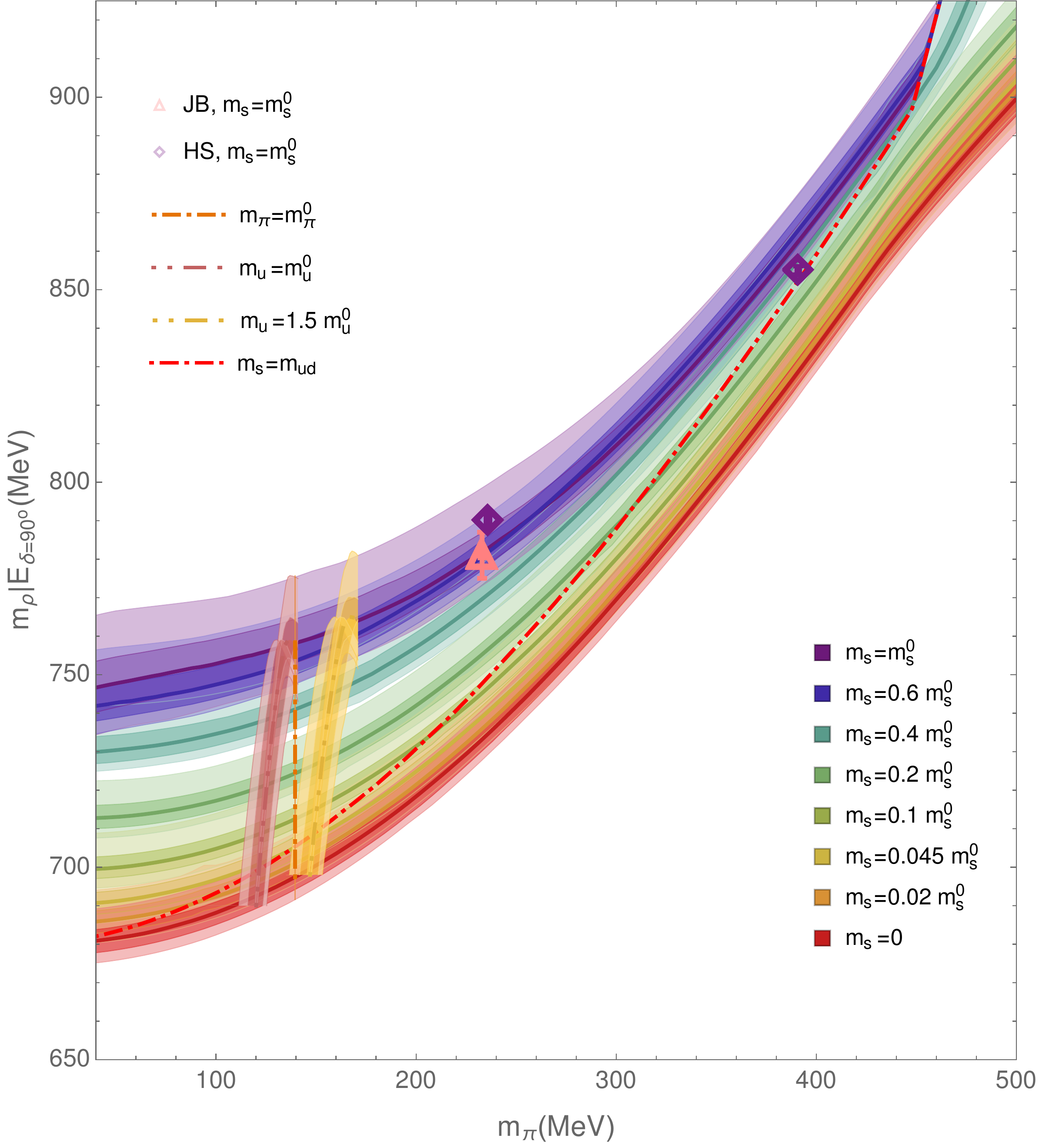} &\includegraphics[scale=0.29]{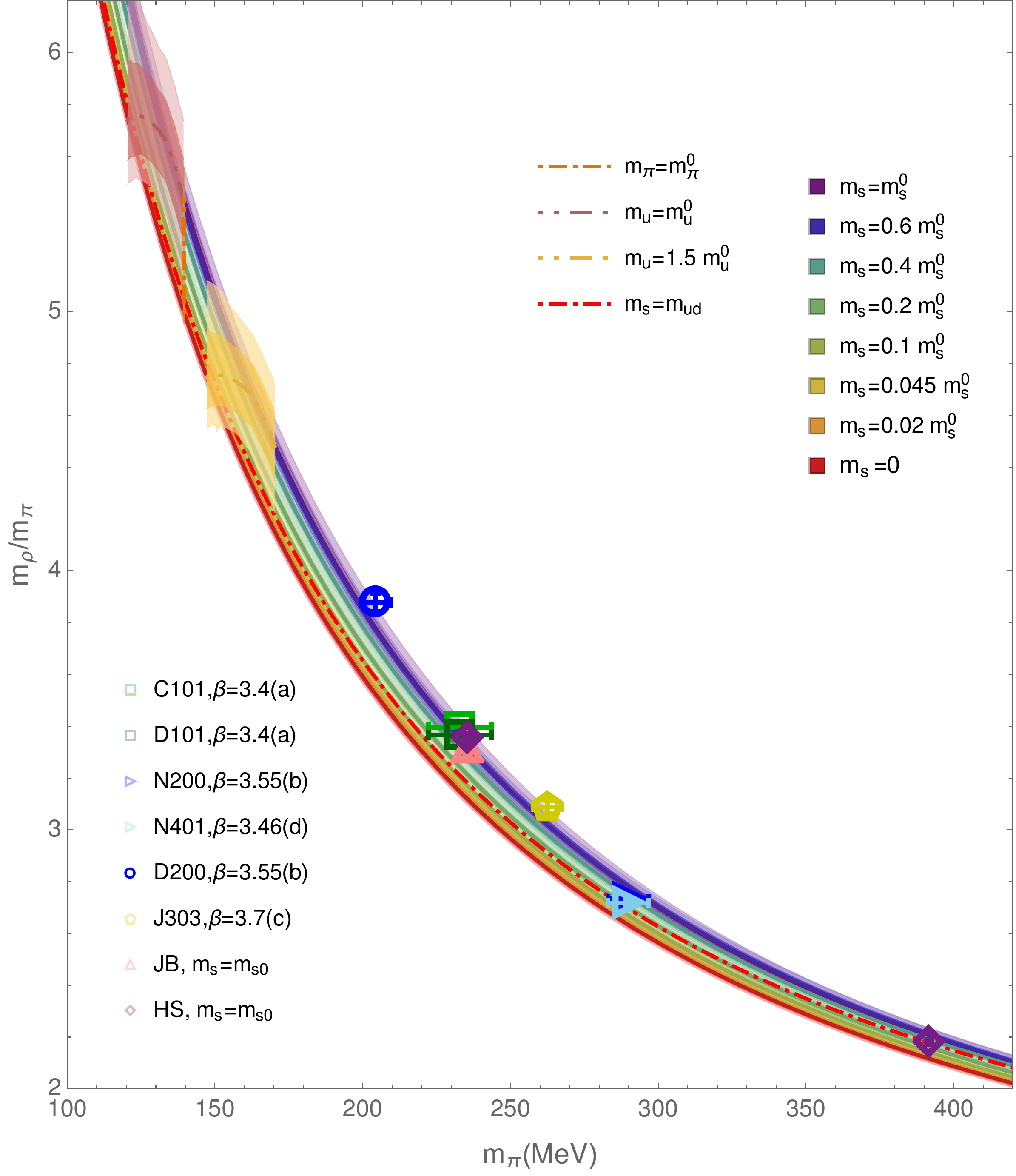}\\
\includegraphics[scale=0.3]{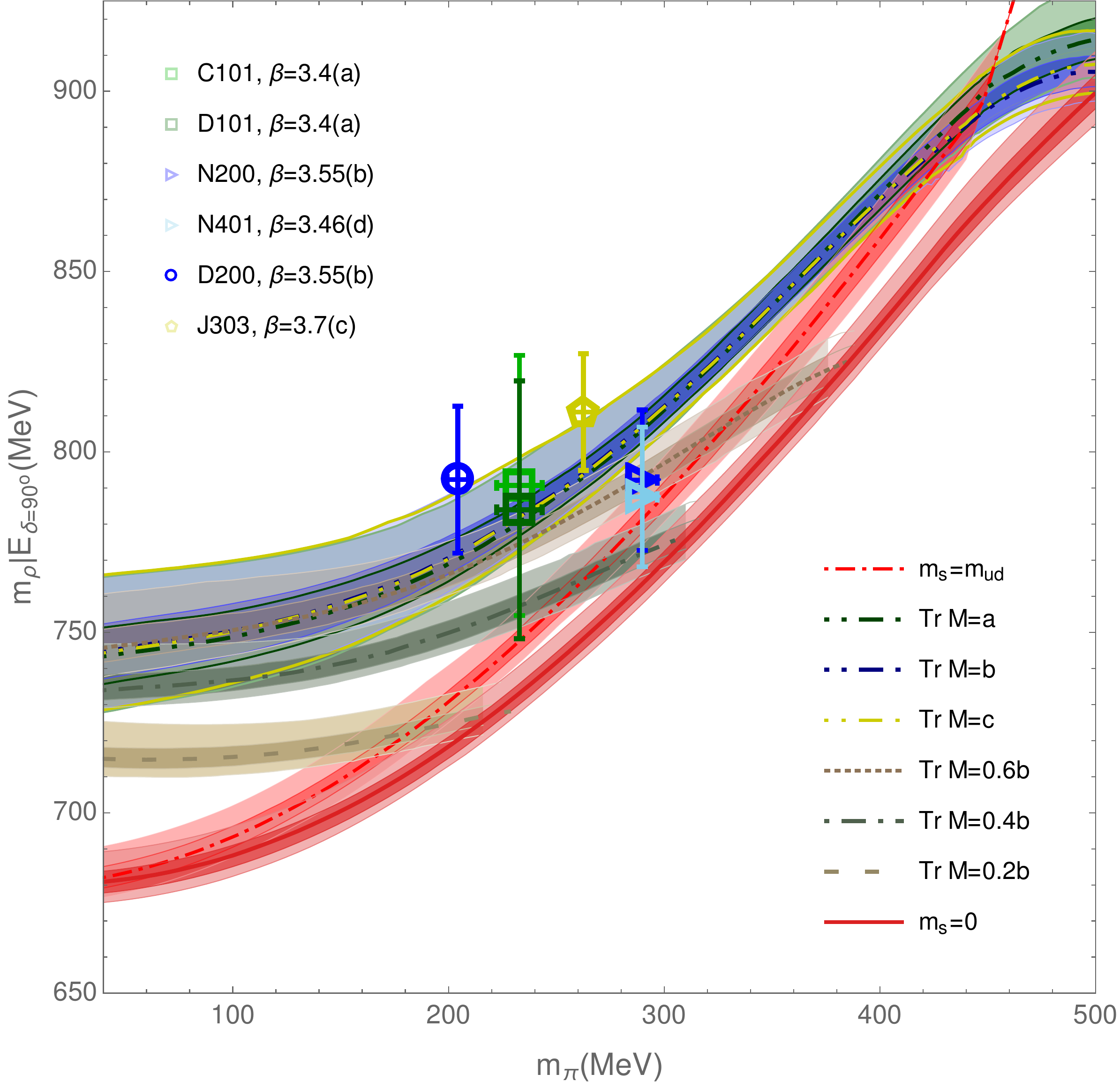}&\includegraphics[scale=0.29]{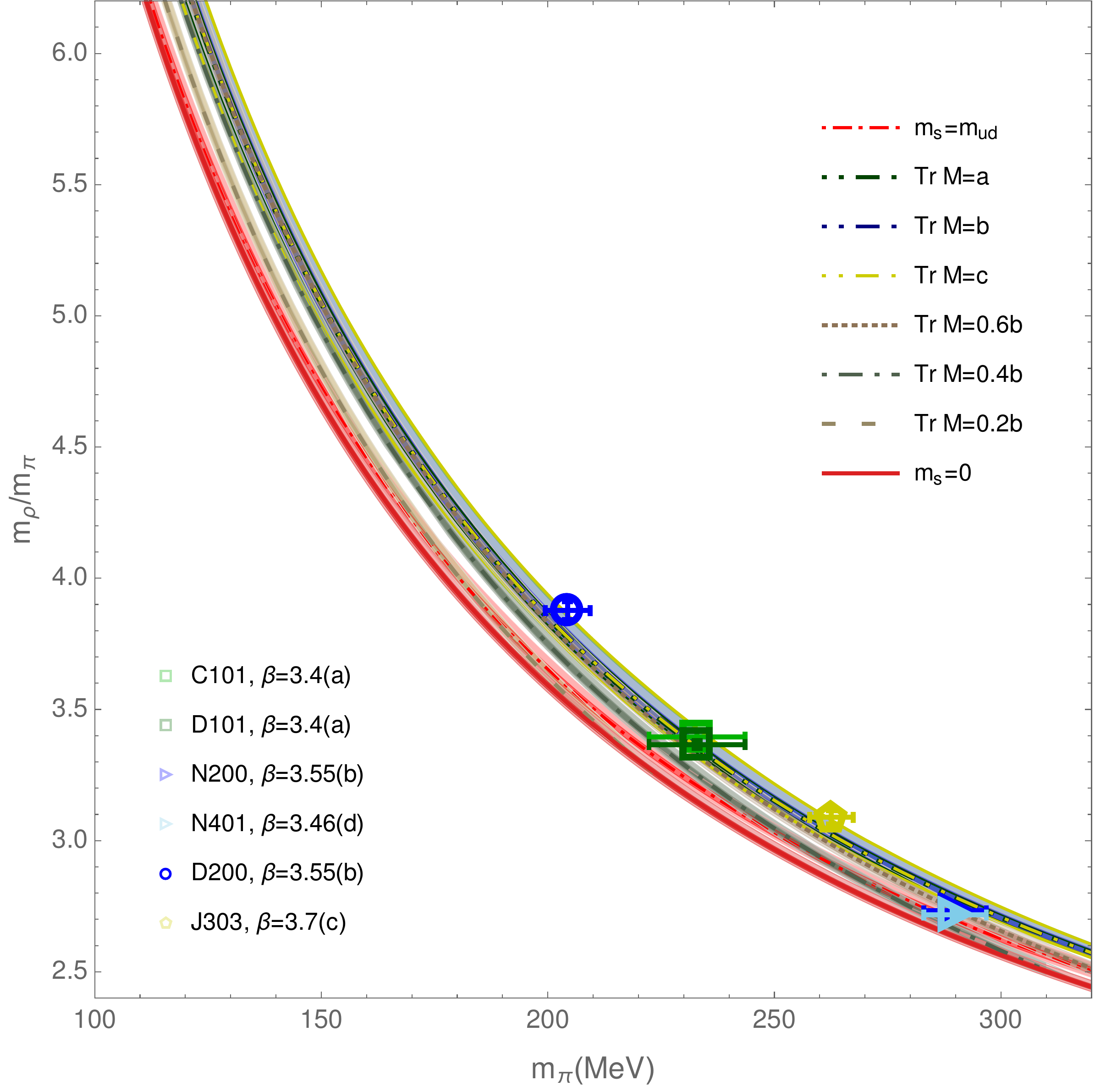}
 \end{tabular}
 \end{center}
\caption{Top: $\rho$-meson mass (left) and this in pion mass units (right) as a function of the pion mass for $m_s=k$, $m_u=r$, $m_\pi=m_\pi^0$, and $m_s=m_u$ trajectories. Bottom: the same for the TrM$=c$ trajectories in comparison with that for the $m_s=0$ and $m_s=m_u$ trajectories, and lattice data.}
\label{fig:mrhoms}
\end{figure*}

In Fig. \ref{fig:mrhoms}, the behavior of the rho meson mass obtained as  $E(\delta=90^0)$ over the $m_s=k$ and TrM$=c$ trajectories is depicted. Indeed, this dependence with the pion mass over the $m_s=m^0_s$ and TrM$=$TrM$^0$ trajectories is indistinguishable till around $400$ MeV. For higher pion masses, the resonance becomes a bound state in the $m_s^0$ trajectory\footnote{When this happens, the $\pi\pi$ threshold is plotted.},
while the $\rho$ gets in between the two thresholds starting to decay in $K\bar{K}$ in the TrM$=$TrM$^0$ trajectory. For other trajectories $m_s=k$ and around the physical point, almost no change in the $\rho$ meson mass is observed. However, when $m_s$ starts to decrease below $0.5\,m_s$, $m_\rho$ decreases faster till around $690-700$ MeV when it reaches the $m_s=0$ line. This value is close to the extrapolation to the physical point obtained from two-flavor lattice data analyses \cite{Hu:2016shf,Guo:2016zos}.
 This is more clear in the $m_{u,\pi}=r$ trajectories, where the mass of the $u$ quark (or pion) is kept fixed. The mass of the $\rho$ meson decreases faster as $m_s$ decreases and the $\rho$ meson starts to decay into $K\bar{K}$. While other TrM$=c$ trajectories present less dependence with the pion mass being flatter.

\begin{figure*}
\begin{center}
\begin{tabular}{cc}
\includegraphics[scale=0.25]{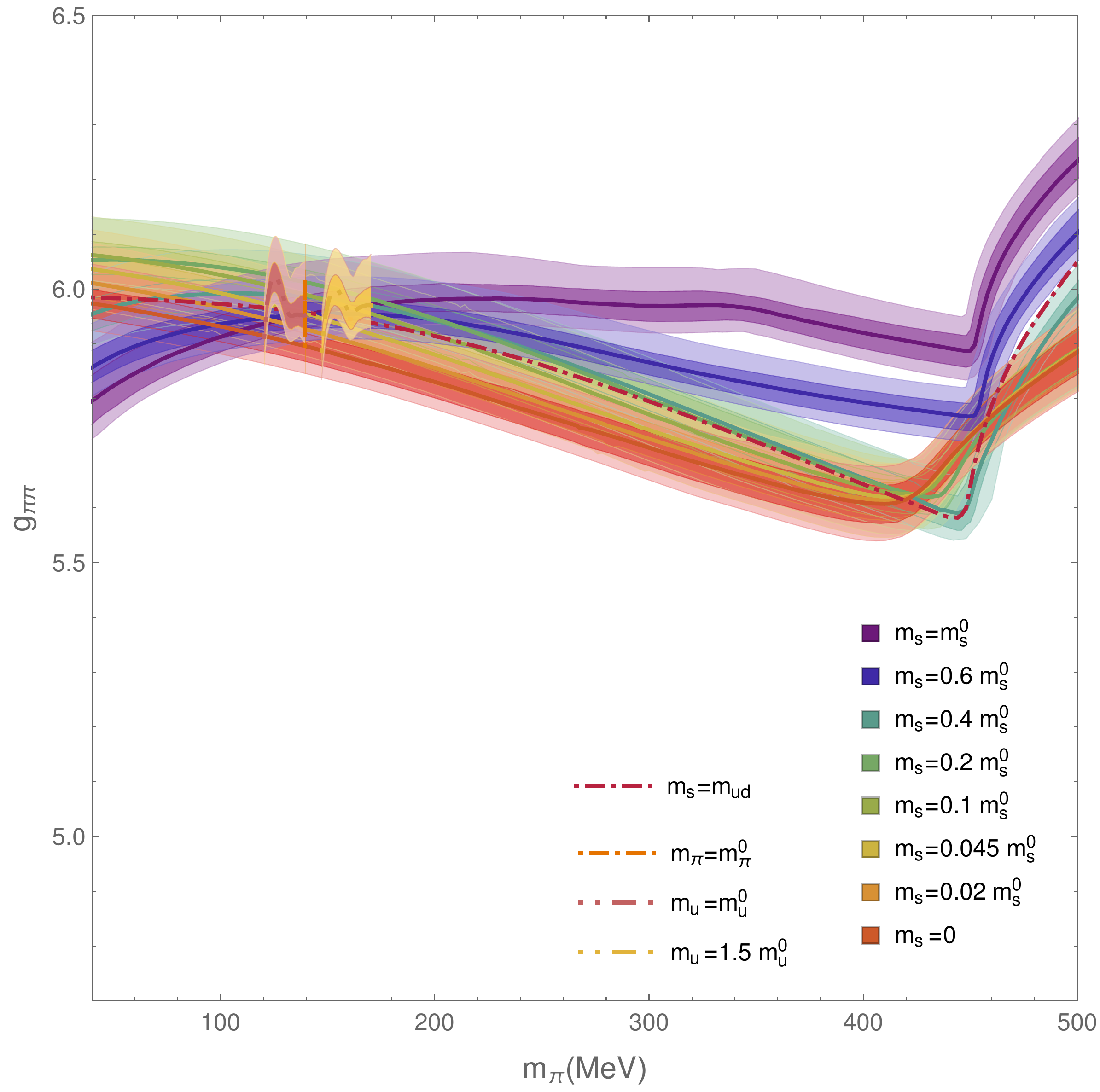} &\includegraphics[scale=0.31]{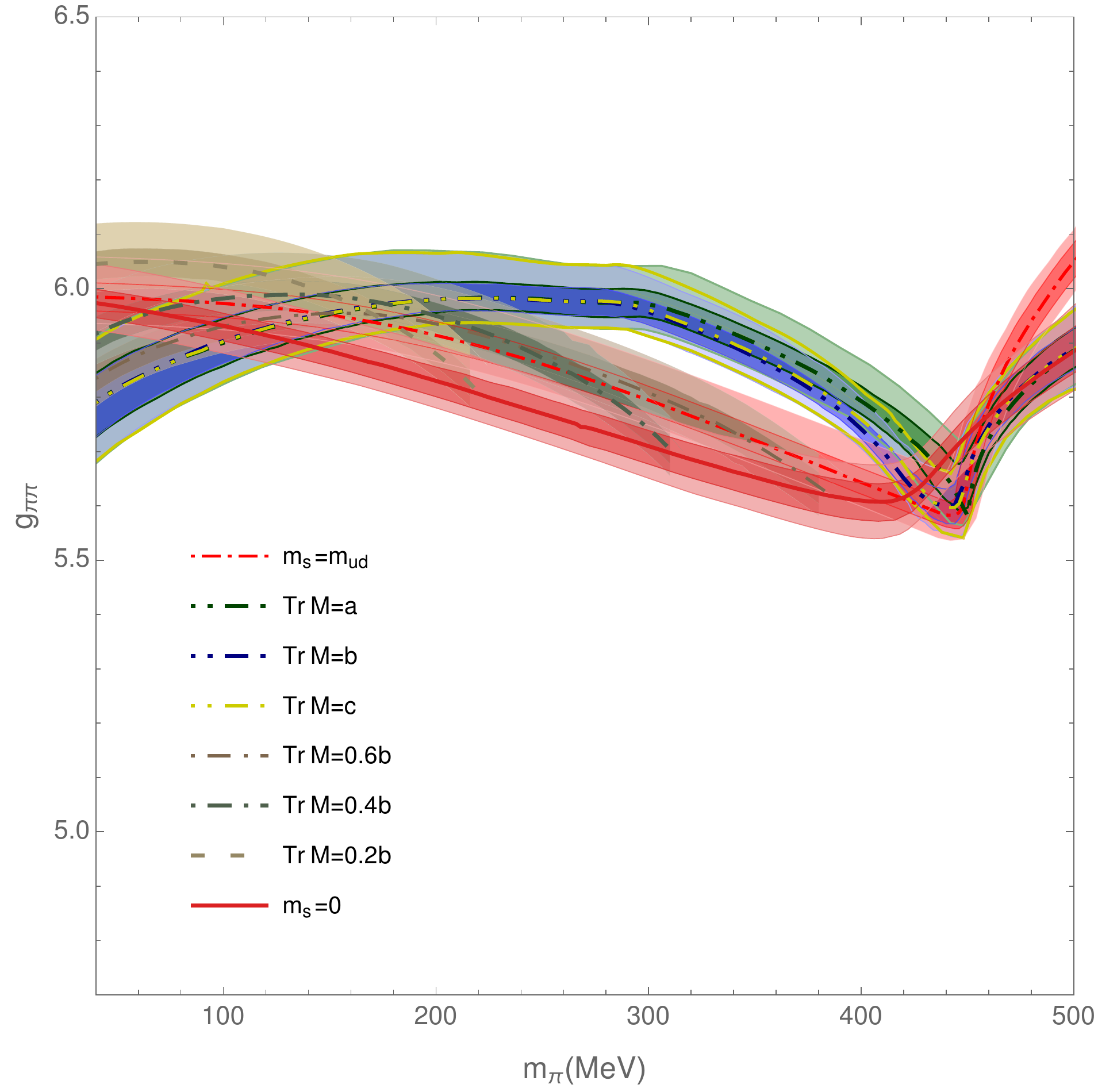}\\
\includegraphics[scale=0.31]{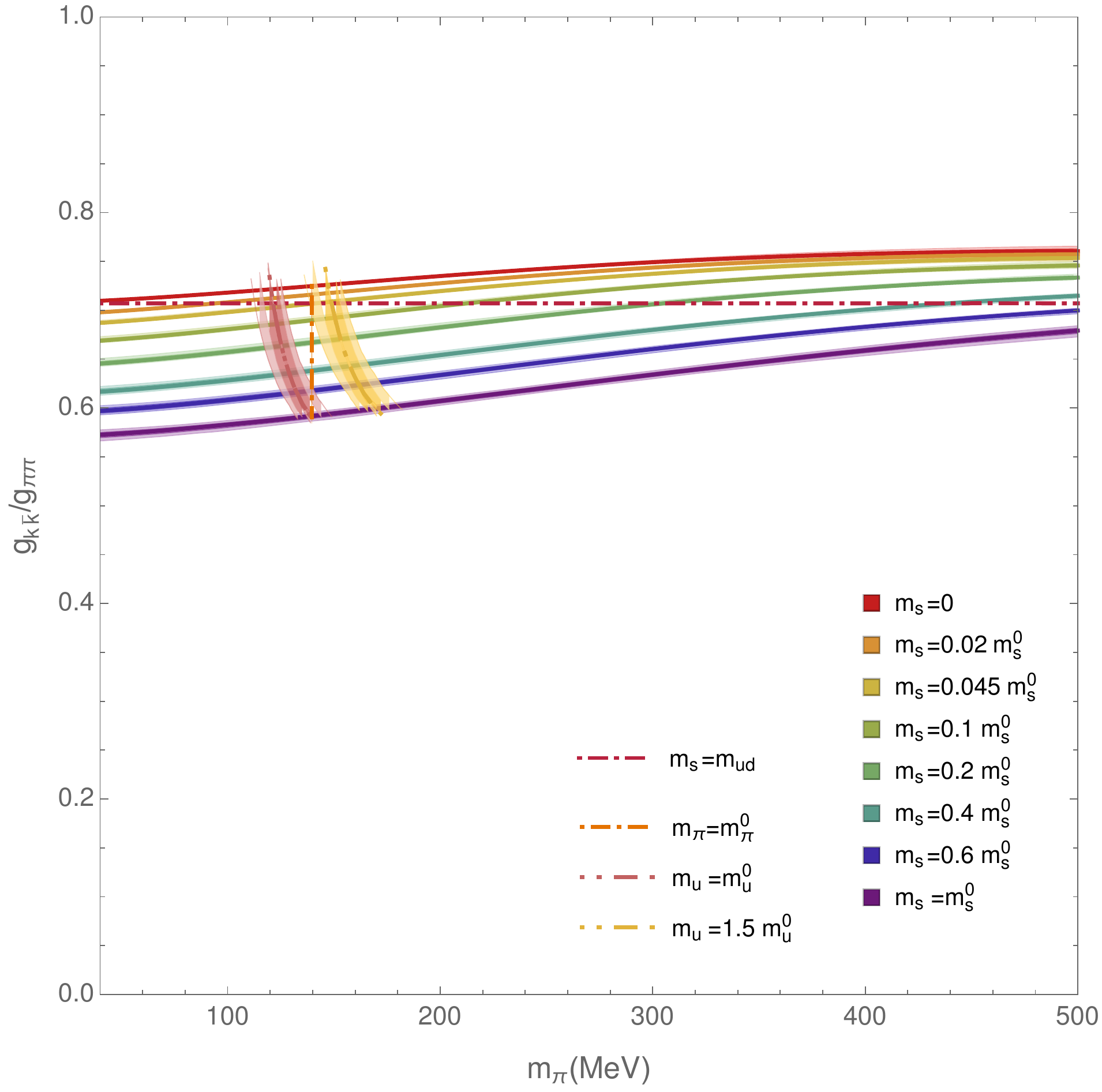} &\includegraphics[scale=0.31]{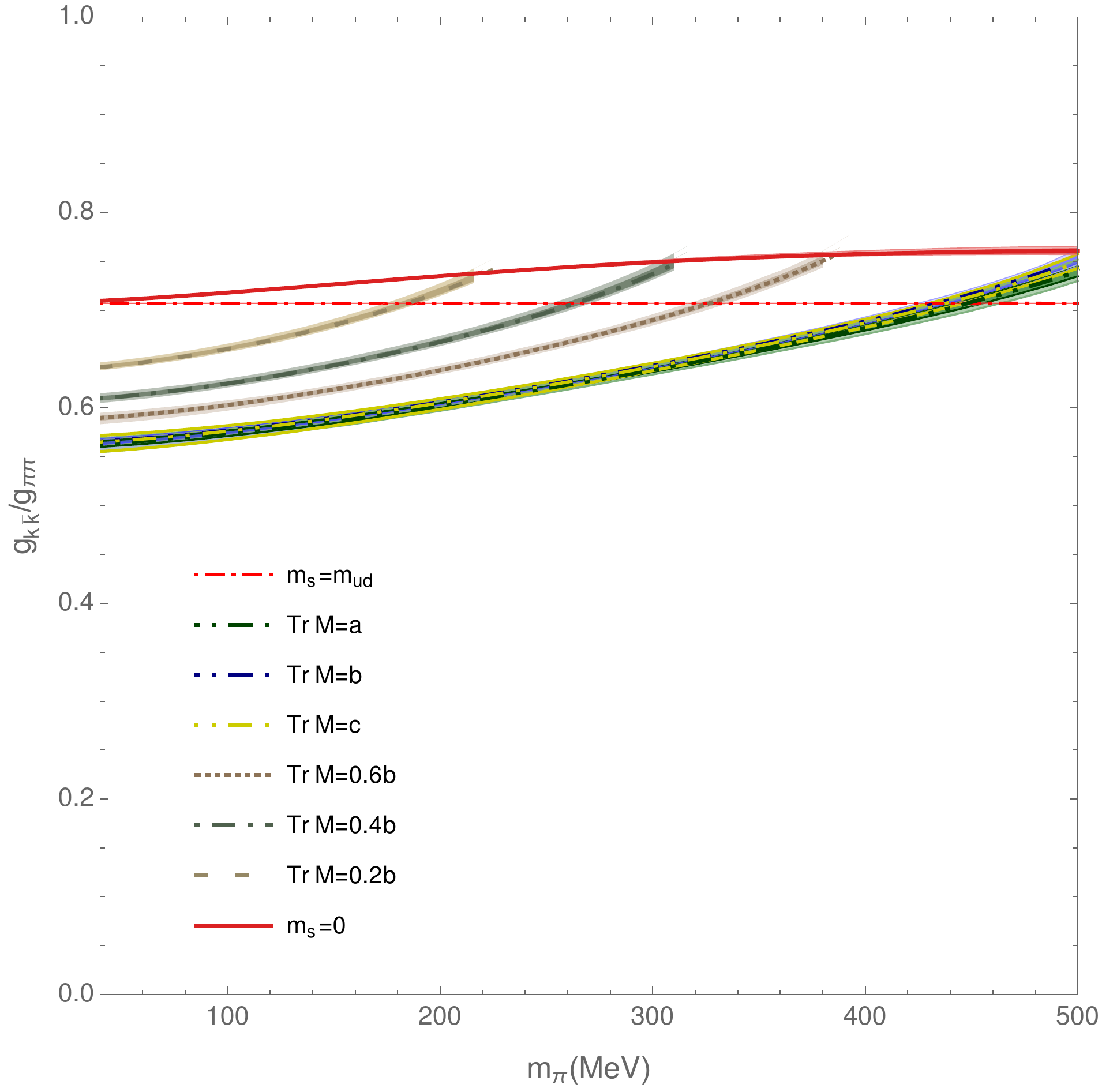}\\
 \end{tabular}
 \end{center}
\caption{Top: couplings of the rho meson to the $\pi\pi$ channel in different chiral trajectories; Bottom: the ratio $g_{K\bar{K}}/g_{\pi\pi}$ in different chiral trajectories.}
\label{fig:gpi}
\end{figure*}

\begin{table}
\begin{center}
{\renewcommand{\arraystretch}{2}
\setlength\tabcolsep{0.15cm}
\begin{tabular}{cccc}
\hline
 $m_K^0/m_\pi^0$&$m_\pi^0/f_\pi^0$&$m_K^0/f_\pi^0$&$m_K^0/f_K^0$\\\hline
 $3.55_{-0.02 (0.05)}^{+0.02(0.04)}$&$1.51 _ {-0.012 (0.03)}^{+0.013 (0.03)}$&$5.33_{-0.05(0.13)}^{+0.05(0.11)}$&$4.45 _ {-0.03 (0.10)}^{+0.04 (0.09)}$\\
 \hline
 \end{tabular}}
\end{center}
\caption{Values of the ratios of pseudoscalar masses and decay constants extrapolated at the physical point as a result of fit which can be interpreted in terms of probability. The central value represents the median (or 0.5 quantile), the first upper and down indices gives the $68$\% CI, while the sum of the absolute values of the two upper(down) indices provide the upper(down) limits of the $95$\% CI.}
\label{tab:physdecayra}
\end{table}

\begin{table}
\begin{center}
{\renewcommand{\arraystretch}{1.7}
\setlength\tabcolsep{0.14cm}
 \begin{tabular}{cccccc}
 \hline
 \multicolumn{6}{c}{$L_i$'s$\times 10^{3}$, $cB_0\times 10^{-3}(\mathrm{MeV}^2)$, $kB_0\times 10^{-3}(\mathrm{MeV}^2)$}\\\hline
  $L_{12}^r$&$L_3^r$&$L_4^r$&$L_5^r$&$L_6^r$&$L_7^r$\\
  $0.36^{+0.02(0.06)}_{-0.02(0.02)}$&$-3.44^{+0.04(0.07)}_{-0.04(0.06)}$&$-0.08^{+0.03(0.05)}_{-0.04(0.03)}$&$0.98^{+0.07(0.06)}_{-0.05(0.04)}$&$0.24^{+0.08(0.16)}_{-0.06(0.05)}$&$0.008^{+0.09(0.12)}_{-0.14(0.15)}$\\
  \end{tabular}}
  {\renewcommand{\arraystretch}{1.7}
\setlength\tabcolsep{0.7cm}
  \begin{tabular}{ccccc}
  $L_8^r$&$c_{\beta=3.4}B_0$&$c_{\beta=3.55}B_0$&$c_{\beta=3.7}B_0$&$kB_0$\\
  $0.098^{+0.10(0.11)}_{-0.11(0.16)}$&$268^{+14(8)}_{-18(20)}$&$254^{+11(7)}_{-18(18)}$&$257^{+12(7)}_{-17(19)}$&$224^{+14(10)}_{-18(20)}$\\
  \hline
 \end{tabular}}\end{center}
 \caption{Values of the parameters obtained in the fit which can be interpreted in terms of probability. The central value represents the median (or 0.5 quantile), the first upper and down indices gives the $68$\% CI, while the sum of the absolute values of the two upper(down) indices provide the upper(down) limits of the $95$\% CI. }
 \label{tab:lecsgl}
 \end{table}
 
Couplings are shown in Fig. \ref{fig:gpi}. In these kind of trajectories, when the $\rho$ approaches the $K\bar{K}$ threshold, its coupling increases, while the coupling to $\pi\pi$ looks quite constant overall. At the symmetric line, the ratio of couplings $g_{\pi\pi}/g_{K\bar{K}}$ is exactly $\sqrt{2}$, what coincides with the ratio of SU(3) Clebsh-Gordan-Coefficients (CGC).

\section{Conclusions}
We performed a global analysis of most recent data on $m_s=m^0_s$, Tr M$=c$ trajectories, including both, phase shifts and decay constant data. The bootstrap method employed here (resampling both energies and lattice spacing) provides a satisfactory solution at 95 \% confidence level. The IAM method has also proven itself to explain the behavior of the $\rho$ meson with variations of the quark masses. The values of LECs obtained can describe both, the $m_s$ and $m_{ud}$ dependence in the $I=1,J=1$ two-(pseudoscalar) meson scattering, being thus, more precise than previous determinations based on the $m_s=m_s^0$ trajectory.
Beyond that, we observed interesting effects which involve the $K\bar{K}$ channel. First, as $m_\pi$ increases and the $\rho$ meson pole gets closer to the $K\bar{K}$ threshold, $g_{K\bar{K}}$ becomes larger, the $\rho$ becomes bound in the $m_s=m_s^0$ trajectories, and starts to decay in $K\bar{K}$ in the TrM$=c$ ones. Second, as $m_s$ decreases, the mass of the $\rho$ starts decreasing faster as it gets closer to the $m_s=0$ line. While, in the symmetric trajectory, we find $g_{\pi\pi}/g_{K\bar{K}}=\sqrt{2}$, corresponding to the SU(3) CGC. Our analysis shows that all operators which could be relevant in the energy region should be considered in the lattice simulation because of the dynamics of the interaction with the quark mass. We hope that the results obtained here motivate the lattice community forward to investigate more on these chiral trajectories, which indeed provide useful information to push forward the field.
\section{Acknowledgements}
 We acknowledge helpful discussions with C. Bernard,
J. Bulava, R. Briceño, M. Bruno, J. Dudek, M. Niehus, S.
Schaefer and to the MILC Collaboration. We also thank
J. Bulava for providing the CLS phase-shift lattice data.
R.M. acknowledges financial support from the Fundacão
de amparo à pesquisa do estado de São Paulo (FAPESP).
JRE is supported by the Swiss National Science Founda-
tion, project No. PZ00P2 174228.

\end{document}